\documentclass[12pt,preprint]{aastex}

\usepackage{psfig}
\usepackage{natbib}

\begin{document}

\title{Energetics of Gamma Ray Bursts}
\author{ Raul Jimenez\altaffilmark{1} David Band\altaffilmark{2}
   and Tsvi Piran\altaffilmark{3}}

\altaffiltext{1}{Department of Physics and Astronomy, Rutgers University, 136
Frelinghuysen Road, Piscataway, NJ08854--8019 USA. 
(raulj@physics.rutgers.edu)}
\altaffiltext{2}{X-2, Los Alamos National Laboratory, Los Alamos, NM,
USA 87545. (dband@lanl.gov)}
\altaffiltext{3}{Racah Institute of Physics, The Hebrew University,
Jerusalem, Israel. (tsvi@nikki.fiz.huji.ac.il)}

\begin{abstract}
We determine the distribution of total energy emitted by gamma-ray
bursts for bursts with fluences and distance information.  Our
core sample consists of eight bursts with BATSE spectra and
spectroscopic redshifts.  We extend this sample by adding four
bursts with BATSE spectra and host galaxy R magnitudes.  From
these R magnitudes we calculate a redshift probability
distribution; this method requires a model of the host galaxy
population.  From a sample of ten bursts with both spectroscopic
redshifts and host galaxy R magnitudes (some do not have BATSE
spectra) we find that the burst rate is proportional to the galaxy
luminosity at the epoch of the burst. Assuming that the total
energy emitted has a log-normal distribution, we find that the
average emitted energy (assumed to be radiated isotropically) is
$\langle E_{\gamma iso} \rangle = 1.3^{+1.2}_{-1.0} \times
10^{53}$ ergs (for H$_0$ = 65 km s$^{-1}$ Mpc$^{-1}$,
$\Omega_m=0.3$ and $\Omega_\Lambda=0.7$); the distribution has a
logarithmic width of $\sigma_\gamma=1.7^{+0.7}_{-0.3}$. The
corresponding distribution of X-ray afterglow energy (for seven
bursts) has $\langle E_{X iso} \rangle = 4.0^{+1.6}_{-1.8} \times
10^{51}$ergs and $\sigma_X = 1.3^{+0.4}_{-0.3}$. For
completeness, we also provide spectral fits for all bursts with
BATSE spectra for which there were afterglow searches.
\end{abstract}


\section{Introduction}

The recent breakthrough in our understanding of gamma-ray bursts resulted from
associating of these events with host galaxies. This association demonstrates
that most, and probably all, bursts are at cosmological distances.  Whether
measured from absorption lines in the continuum of optical transients or
emission lines in the host galaxies' spectra, the redshifts show that the
bursts are even further than predicted by the simplest ``minimal''
cosmological model \citep{BandHartmann98}.  With assumptions about the
radiation pattern, the energy radiated by the burst is calculated from the
redshifts and the burst fluence.  The position of the burst within the host
argues for a strong connection between star formation and the burst
\citep{Bloom01}.  Many of the galaxies show evidence of vigorous star
formation.

We now have a sufficient sample of detected host galaxies to consider various
distributions of burst properties which require the distance to the burst.
Here we calculate the distributions of the total energy and peak luminosity
radiated by the burst and the X-ray energy in the afterglow. The energy scale
is a strong constraint on burst scenarios; some proposed sources are
insufficient to provide the necessary observed energy.  However, the width and
shape of the energy distributions are also consequences of the emission
process.  For example, in the current physical scenario
\citep{NPP92,PX94,RM94,SP97} the gamma-rays are radiated by ``internal''
shocks resulting from the collision of regions with different Lorentz factors
within a relativistic outflow. Thus the total gamma-ray emission may depend on
the vagaries of the central source accelerating the outflow.  However, the
afterglow is attributed to the ``external'' shock where the outflow collides
with the external medium.  Thus the afterglow may be a bolometric measurement
of the energy content of the relativistic flow.  This is the basis of the
``patch shell'' model's prediction that the gamma-ray energy distribution
should be broader than that of the afterglows \citep{KP99}.

To expand the burst sample, we include not only bursts with spectroscopic
redshifts, but also bursts for which there are only host galaxy magnitudes.  A
galaxy magnitude can be mapped into a probability distribution for the burst's
redshift.  This mapping requires a model of the host galaxy distribution;
competing models are tested by comparing the measured spectroscopic redshifts
with the magnitude-derived redshift distribution for those bursts with both a
spectroscopic redshift and a host galaxy magnitude.  This host galaxy model is
intrinsically interesting since it indicates whether the burst rate is
proportional to the host galaxy mass or luminosity.

An underlying assumption in our study is that the burst energy and
luminosity distributions have not changed over the universe's
lifetime.  Similarly, we assume that a burst's energy or
luminosity is uncorrelated with the host galaxy magnitude.

In \S 2 we present the expected fluence distribution given the
assumed lognormal burst energy distribution.  \S 3 describes the
sources of the observed $\gamma$-ray and X-ray energy fluxes and
fluences. \S 4 discusses the determination of a burst redshift
from a single optical magnitude of the host galaxy. However,
galaxy surveys count individual galaxies while the burst rate may
be proportional to the mass or luminosity of a galaxy; therefore
in \S 5 we consider the underlying host galaxy luminosity
function. The burst $\gamma$-ray and X-ray energy distributions
are calculated in \S 6.  Our conclusions are summarized in \S 7.

\section{The Expected Fluence Distribution}

Here we present the methodology for modeling the distribution of
the total burst energy $E_\gamma$ given a set of observed burst
fluences. The determination of the distribution of
the peak gamma-ray luminosity and of the X-ray afterglow energy
are analogous.

We begin by assuming that $E_\gamma$ has a log-normal
distribution,
\begin{equation}
p(E_\gamma \,|\, \langle E_{\gamma iso} \rangle,\sigma_\gamma)
   d(\log E_\gamma)
   = {1\over{\sqrt{2\pi}\sigma_\gamma}}
   \exp \left[-{{(\log(\langle E_{\gamma iso} \rangle)-\log(E_\gamma))^2}
   \over{2\sigma_\gamma^2}} \right] d(\log E_\gamma)
   \quad .
\end{equation}
Thus the bolometric fluence $F=E_\gamma (1+z)/4\pi D_L(z)^2$
(where $D_L(z)$ is the luminosity distance) has the log-normal
distribution
\begin{equation}
p(F \,|\, \langle E_{\gamma iso} \rangle,\sigma_\gamma) d(\log F)
   = {1\over{\sqrt{2\pi}\sigma_\gamma}}
   \exp \left[-{{(\log(\langle E_{\gamma iso} \rangle
   (1+z)/4\pi D_L(z)^2)-\log(F))^2}
   \over{2\sigma_\gamma^2}}\right] d(\log F)
   \quad .
\end{equation}
Note that $\sigma_\gamma$ is a width in logarithmic space, and the
linear change of variables does not affect this width.  This
probability for the fluence assumes that the redshift is known. If
not, then we have to convolve eq.~2 with the probability
distribution for the host galaxy redshift, $p_B(z\,|\,D_o)$, where
$D_o$ is the optical data we have about the redshift (e.g., the
measured spectroscopic redshift from spectral lines or the host
galaxy magnitude).  The resulting probability is
\begin{eqnarray}
&& p(F \,|\, \langle E_{\gamma iso} \rangle,\sigma_\gamma)
   d(\log F) = \\
&&\int dz\,p_B(z\,|\,D_o)
{1\over{\sqrt{2\pi}\sigma_\gamma}}
   \exp \left[-{{(\log(\langle E_{\gamma iso} \rangle (1+z)/4\pi
   D_L(z)^2)-\log(F))^2}
   \over{2\sigma_\gamma^2}} \right] d(\log F) \quad . \nonumber
\end{eqnarray}

These are the distributions for $E_\gamma$ and $F$, regardless of
whether these quantities are actually observable.  The
distributions for the observable bursts must therefore be
truncated at the threshold value of the fluence, $F_T$, and the
overall distribution must be renormalized.  Thus
\begin{eqnarray}
p_{\hbox{obs}}(F \,|\, \langle E_{\gamma iso}\rangle
   ,\sigma_\gamma) =
   {{p(F \,|\, \langle E_{\gamma iso} \rangle,\sigma_\gamma)\theta(F-F_T)}
   \over{\int_{F_T}^\infty dF \,
   p(F \,|\, \langle E_{\gamma iso} \rangle,\sigma_\gamma)\theta(F-F_T)}}
\end{eqnarray}
where $\theta(x)$ is the Heaviside function (1 above $x=0$, and 0
below).  Since BATSE triggers on the peak count rate
$C_{\hbox{max}}$ and not the fluence, the threshold fluence $F_T$
is the observed fluence divided by
$C_{\hbox{max}}/C_{\hbox{min}}$, where $C_{\hbox{min}}$ is the
threshold count rate (i.e., the minimum peak count rate at which
BATSE would have triggered). Thus $F_T$ and $p_{\hbox{obs}} (F
\,|\, \langle E_{\gamma iso} \rangle, \sigma_\gamma)$ vary from
burst to burst.  For calculations of the X-ray afterglow energy
the X-ray detectability threshold must be used instead of $F_T$.
The likelihood is the product of these probabilities evaluated
with each burst's observed properties (fluence, redshift, etc.)
for all bursts in the sample. By maximizing this likelihood we
find the preferred values and confidence ranges for $\langle
E_{\gamma iso} \rangle$ and $\sigma_\gamma$ (or the equivalent for
the X-ray afterglow).  Note that the selection effect introduced
by observing bursts at different redshifts---only intrinsically
bright high redshift bursts are observable---while the detector
threshold varies is mitigated by truncating the probability
distribution for the observable fluence at the threshold value.

\section{$\gamma$-Ray and X-Ray Data}

We consider the sample of GRBs with observed optical afterglows
and with detected host galaxies, whether or not a redshift has
been measured, for bursts between 1997 January and 2000 June ({\it
CGRO}'s untimely demise). To obtain a uniform estimate of the
$\gamma$-ray energy fluence we consider only those bursts detected
by BATSE.  The bursts with BATSE data and a spectroscopic redshift
are GB970508, GB970828, GB971214, GB980425, GB980703, GB990123,
GB990506, GB990510, and GB991216; however, GB980425 is exceptional
because of its possible association with a supernova in a
relatively nearby galaxy and we do not include it in our sample.
The bursts with BATSE data and only an R magnitude for its host
galaxy are GB971227, GB980326, GB980329, and GB980519.   In
addition, both R magnitudes and spectroscopic redshifts but no
BATSE data are available for GB970228, GB980613, and GB990712. For
GB990510 there is currently a spectroscopic redshift but no host
galaxy R magnitude.  Thus, there are eight bursts for which we can
calculate the total emitted energy, and another four for which we
can calculate a range of possible energies.  There are ten bursts
for which we have both spectroscopic redshifts and a host galaxy R
magnitude; from these ten we determine a model of the host galaxy
population, as described below.

To calculate the total gamma-ray energy emitted by a burst we need
the observed gamma-ray fluence.  We determine the fluence of a
given burst by fitting the spectrum accumulated over the time
segment during which there was detectable emission.  For strong
bursts we use spectra from BATSE's Spectroscopy Detectors (SDs),
while for weak bursts we fit spectra from BATSE's Large Area
Detectors (LADs).  Since BATSE consisted of 8 modules, each with
an SD and an LAD, more than one detector observed each burst.  For
strong bursts we choose the SD with the smallest burst angle (the
angle between its normal and the burst) and with a gain
sufficiently high to cover the 30--1000~keV energy band (the SDs
were operated at different gains).  A typical SD spectrum provides
$\sim 200$ usable energy channels.  For weak bursts we use the LAD
with the highest count rate which almost always had the smallest
burst angle.

\begin{figure}
\plotone{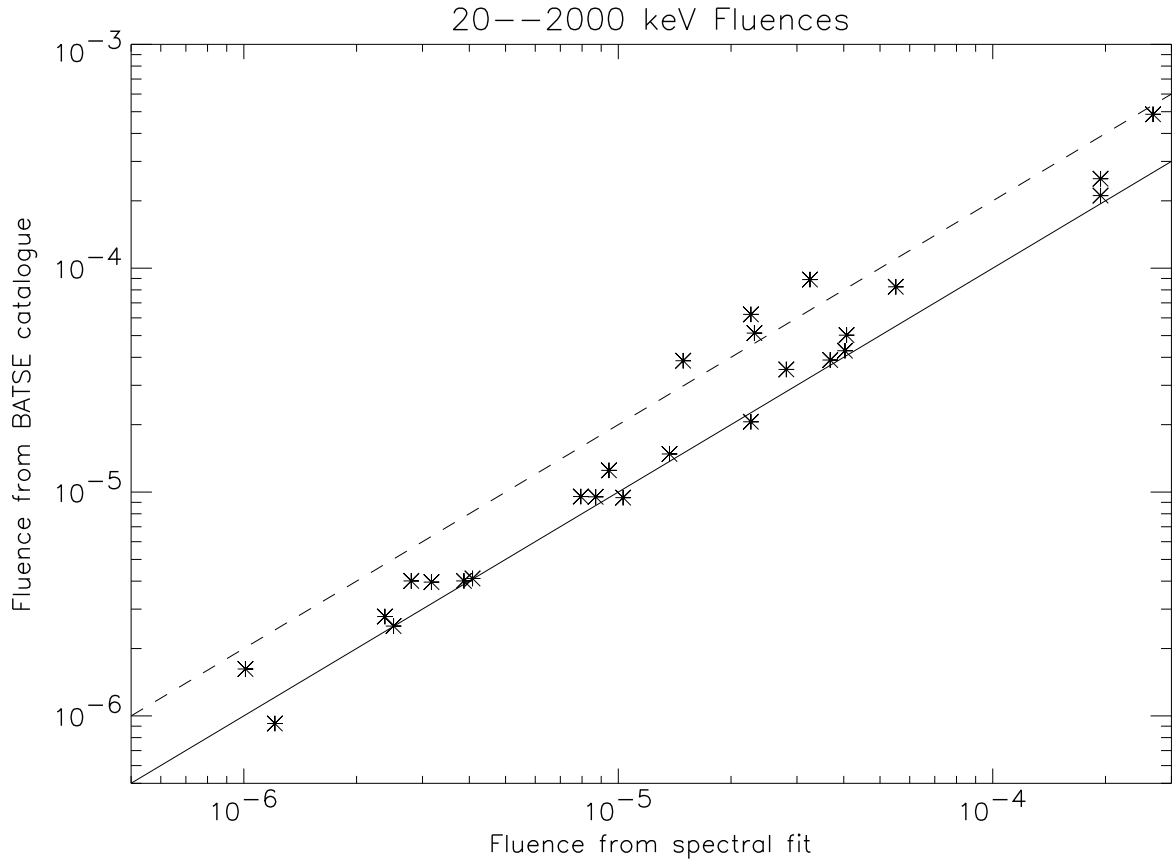}
\caption{
Comparison between the 20--2000~keV
fluences (in erg cm$^{-2}$) from the BATSE catalog and from
integrating the spectral fits in the table over energy and time.
The fluences are equal along the solid line while along the dashed
line the catalog fluences are twice as great as the fluences
derived from the spectral fits.\label{flu_comp}}
\end{figure}

These spectra were fitted with the ``Band function''
\cite{Band+93}, which consists of a low energy power law cut off
by an exponential, $N_E=N_0 (E/100\hbox{ keV})^\alpha
\exp[-E/E_0]$ ph s$^{-1}$ keV$^{-1}$ cm$^{-2}$, which merges
smoothly with a high energy power law, $N_E\propto E^\beta$.  The
parameters of the resulting fits are given in the 2nd--5th columns
of Table~\ref{Table}.  The energy fluence $F_\gamma$ (6th column
of the table) is obtained by integrating this fit over the energy
band 20--2000~keV and over the time for which there was detectable
emission (the 10th column).  This is the energy band of the
fluences provided by the BATSE catalogue.  In most cases the
fluence in this energy band is very nearly the bolometric fluence,
and thus no k-corrections are necessary to calculate the energies
for bursts at different redshifts (the k-correction compensates
for the redshift shift in the energy band between the emitter and
the detector). In this table we provide for completeness spectral
fits and fluences for all the bursts observed by BATSE which were
localized after 1997 January. Figure~\ref{flu_comp} compares the
20--2000~keV fluences provided by the BATSE catalogues and
calculated here; as can be seen, the BATSE catalogue values are
somewhat larger than the fluences resulting from spectral fits.
Note that there are host galaxy observations for only 12 of the
bursts in this table; these 12 burst comprise our ``BATSE redshift
sample.''

As described above, for each burst we need the threshold fluence,
the minimum fluence at which BATSE would have triggered.  This is
derived by equating the ratios of the observed and threshold
fluences and peak count rates: $F_T = F / [C_{\hbox{max}} /
C_{\hbox{min}}]$.  The online BATSE catalog provides
$C_{\hbox{max}} / C_{\hbox{min}}$ for some bursts, but values are
missing for many bursts because of data gaps.  For bursts without
a catalog value we have estimated $C_{\hbox{max}} /
C_{\hbox{min}}$ from the light curves.

In addition to the energy emitted by the burst (mostly in
gamma-rays), significant emission occurs during the first few
hours of the afterglow (mostly in X-rays).  We estimate this
energy (excluding the prompt X-ray emission during the burst)
using the observed late (few hours) X-ray flux
in the 2--10~keV band and assuming this emission decays as a power
law with the index $\delta_x$ shown in Table~\ref{Table} over the
period 100 to $10^5$ seconds after the burst.

\section{Host Galaxy Redshift Distribution}

The methodology presented in \S 2 requires $p_B(z\,|\,D_o)$, the
probability that the burst redshift has a given value.  Thus far
redshift information has been derived from the associated host
galaxies.  Clearly, if we know the spectroscopic redshift $z_0$
then $p_B(z\,|\,D_o) = \delta(z-z_0)$. However, spectroscopic
redshifts are not available for some of the host galaxies.  For
these host galaxies we derive a redshift probability distribution
with a finite width from their observed single-band magnitudes.

The use of photometry to derive galaxy redshifts has flourished
during the past 5 years (e.g. \citet{CCSKM95, SLY97, D+98, FLY99,
A+99}). Redshifts are now commonly determined to an accuracy of
10\% (e.g., \citet{CCSB99}) using multi-band photometry (usually 4
to 6 bands). Multi-band photometry is therefore an economical
method of determining redshifts. Unfortunately, most GRB hosts
have been observed in only a single band (typically $R$).

It may seem futile to try to determine the redshift of a galaxy
from a single band. Galaxies span a large range in stellar masses
(i.e., luminosities) and therefore galaxies of a given magnitude
are found at virtually any redshift (c.f., \citet{D99}).  On the
other hand, since the volume element $dV/dz$ peaks at around
$z=1-2$ (for reasonable cosmologies) and galaxies formed stars at
a higher rate at $z>1$ (and consequently were brighter than at
present), we expect a high probability that the average galaxy
inhabited the $1<z<2$ region for not too faint magnitudes.  Here
we use the Hubble Deep Field (HDF) to derive a galaxy redshift
distribution at different observed magnitudes. With this
distribution we then determine a probability distribution of the
host galaxies' redshift based on their observed optical
magnitudes.  Our methodology permits us to consider different
redshift distributions corresponding to different models of the
host galaxy population (e.g., the burst rate is proportional to
the galaxy luminosity); we test the hypotheses about the host
galaxies by determining whether the model redshift distributions
are consistent with the host galaxies with spectroscopic
redshifts. We use only observations from the HDF, the deepest
available survey, in order not to complicate our analysis with
theoretical prejudices in merging results from different surveys.
Note that we only require the distribution of redshifts at a given
magnitude, and thus as long as a redshift is determined for every
galaxy at a given magnitude in the survey field, the HDF will
suffice.

\begin{figure}
\plotone{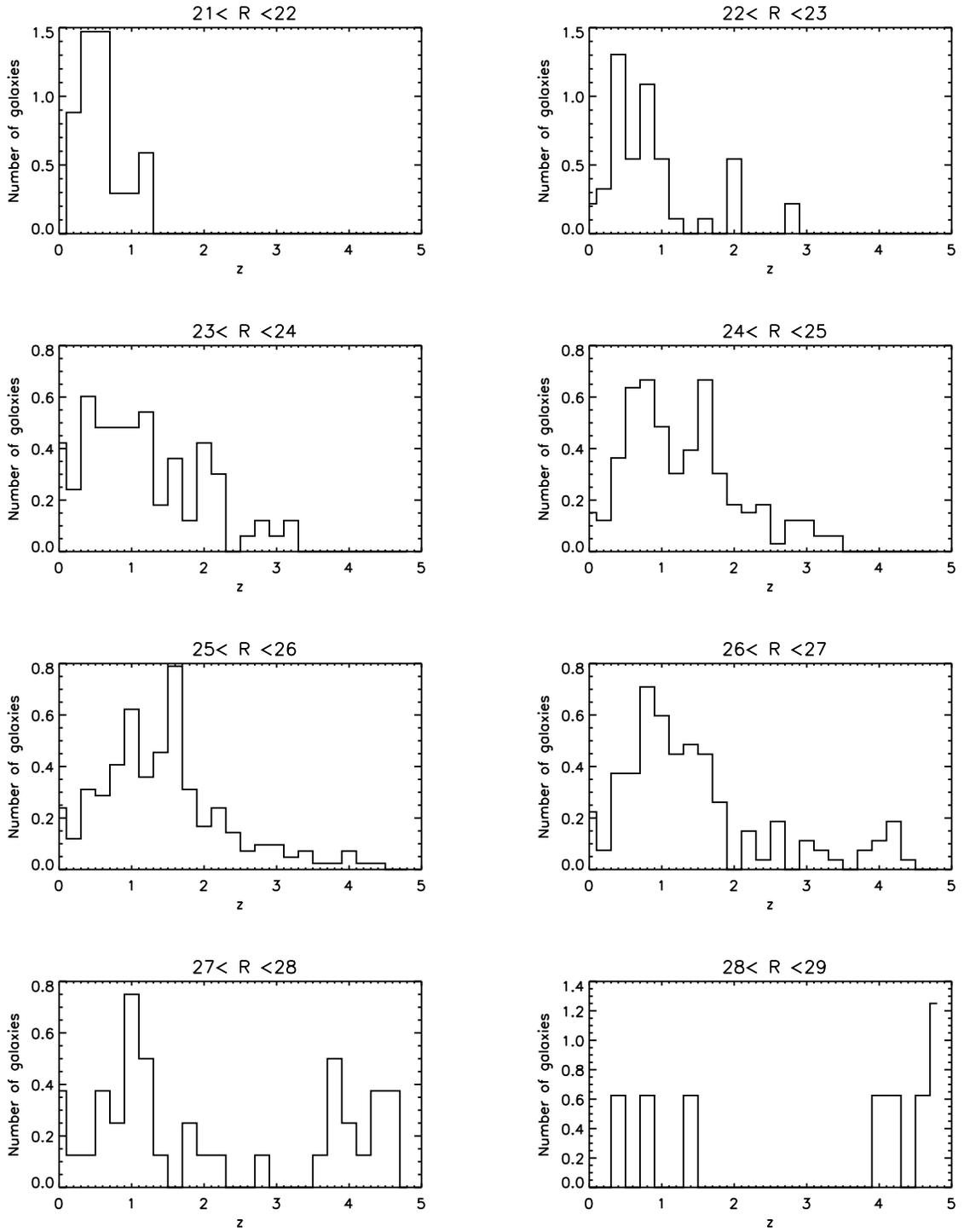}
\caption{
The redshift distribution of galaxies in the $R$ band from the HDF.\label{hdf}} 
\end{figure}

Spectroscopic redshifts have thus far not been determined for all
the HDF galaxies.  However, the available multi-band photometric
redshifts (accurate to 10\%) will suffice.  We have used the HDF
catalog from \citet{D99}, to which the reader is referred for a
full description of the sample and the expected accuracy of the
photometric redshifts. In Fig.~\ref{hdf} we show the galaxy
redshift distribution as a function of apparent $R$ magnitude,
where we have transformed the HST F606W magnitudes into cousins
$R$ magnitudes.  As expected, $R < 26$ galaxies are systematically
more common at $z<2$ than at higher redshifts.

When normalized, the distributions in Fig.~\ref{hdf} provide the
galaxy redshift probability distributions in a given $R$ magnitude
band. Thus the HDF provides $p_{\hbox{HDF}}(z \,|\, f_R)$---$f_R$
is the energy flux corresponding to an R magnitude---from which we
derive $p_B$ for those bursts without a spectroscopic redshift.
Galaxy surveys weight each detected galaxy equally.  However, the
burst rate may not be the same for each galaxy, and thus we have
to weight the galaxies in the HDF based on a model of burst
occurrence.  To determine how to weight $p_{\hbox{HDF}}$ we have
to analyze the origin of this probability: it is the normalized
product of the comoving volume per redshift, $dV/dz$, times the
galaxy density at $L_R$ per comoving volume, $n(L_R,z)$. $L_R$ is
the luminosity over the band
in the burst's frame that redshifted into the $R$ band in the
observer's frame; this band varies with redshift.  Note that we do
not need k-corrections for $L_R$ since the density is for those
galaxies which provide us with the observed $f_R$ in the $R$-band
in our frame. Thus
\begin{equation}
p_{\hbox{HDF}}(z\,|\,f_R) dz =
   {{n(L_R=4\pi D_L^2f_R,z) {{dV}\over {dz}} dz }
   \over{\int n(L_R=4\pi D_L^2f_R,z) {{dV}\over {dz}} dz}}
   \quad .
\end{equation}

We consider four models.  First, we use $p_{\hbox{HDF}}$ as $p_B$;
every galaxy has an equal probability of hosting a burst.  This is
inconsistent with most burst scenarios, but serves as a useful
null hypothesis.

Second, we assume that the burst rate is proportional to the mass
of the galaxy.  This corresponds to a scenario where the burst
occurs long after the progenitor was formed, and thus the
progenitor population is (approximately) proportional to the
galaxy's mass.  In this case $p_B$ is $p_{\hbox{HDF}}$ weighted by
the luminosity the galaxy would have today in the R-band.  Thus
{\it both} a k- and an e-correction are required, where the
e-correction compensates for the evolution in the galaxy's
spectrum. Note that these k- and e-corrections are {\it not}
applied to $p_{\hbox{HDF}}$ but only to the weighting. Thus $p_B
\propto L_R K(z)E(z)p_{\hbox{HDF}}(z \,|\,f_R)$ where $K(z)$ is
the k-correction which references $L_R$ (the luminosity in the
band at the host galaxy's redshift which is redshifted into
today's R-band) to today's R-band and $E(z)$ is the e-correction
which maps the R-band luminosity of a galaxy at the host's
redshift to the R-band luminosity it would have at present.

Third, we assume that the burst rate is proportional to the
galaxy's intrinsic luminosity at the time of the burst.  This
corresponds to scenarios where the burst occurs very soon after
the progenitor forms.  Thus, if the luminosity per mass was
greater in the past as a result of increased star formation, then
the burst rate should also increase.  This is admittedly an
oversimplification of a galaxy's evolutionary history.  For this
model $p_B \propto L_R K(z)p_{\hbox{HDF}}(z \,|\,f_R)$ where, once
again, $K(z)$ is the k-correction which references $L_R$ (the
luminosity in the band which is redshifted into today's R-band) to
today's R-band.

Fourth, we model the star formation rate for a given model by weighting the
second model, where the burst rate is proportional to the galaxy mass, by the
cosmic redshift-dependent star formation rate.  This rate rises rapidly with
redshift to $z\sim 1$, and then levels off, e.g. cite{H+98}.

Note that the $p_B$ for each of these different models needs to be
renormalized given the model-dependent weightings.

Whenever we need to compute a k and/or e--correction in any of the above
scenarios, we have used the set of synthetic stellar populations developed in
\cite{JPMH98}. The star formation rate is modelled as a declining exponential,
with e-folding ($\tau$) time depending on the morphological type. In
particular, we used the following values: $\tau=1$, 3, 5 and $\inf$ for E/S0,
Sab, Scd and Irr types respectively. We kept the metallicity fixed at the
solar value during the whole evolution of the galaxy and chose $z=4$ as the
redshift of formation -- starting point of the star formation -- for all
galaxies. This modeling of galaxies is, indeed, overly simplistic, but it
is sufficient for our purposes since we do not require to be precise by more
than 0.5 magnitudes when computing the k+e corrections.

\section{Testing the Host Galaxy Models}

We can test the different models for the host galaxy population
using the bursts with both spectroscopic redshifts and host galaxy
R magnitudes.  We compare the likelihoods using the various
models.  However, the probability used in these likelihoods is the
probability that we would determine the redshift that was indeed
observed, not the probability that a host galaxy with a given R
magnitude would have a particular redshift.  For the telescope
which determined the redshift there are redshift windows in which
it would not have been able to make this determination.  For
example, for the Keck observations the redshift is difficult to
determine between $z=$1.3, when O[II] is redshifted out of the
spectrum, and $z=$2.5, when Ly$\alpha$ is blue shifted in. Thus in
the likelihood we should use $p\propto h(z)p_B$, where $h(z)$ is
the probability of determining a particular value of the redshift;
however, including the redshift windows applicable to each
observation is beyond the scope of this paper. Note that by
modifying the distribution of host galaxy redshifts, whether
observable or not, for a given R magnitude into the distribution
of host galaxies which are actually observable, we avoid the
selection effect resulting from the easier detectability of bright
host galaxies.

The resulting likelihood is
\begin{equation}
\Lambda = \prod_i {{h(z_i)p_B(z_i \,|\, f_{R,i})}\over
   {\int dz \, h(z_i)p_B(z_i \,|\, f_{R,i})}} \quad .
\end{equation}
We find that $\Lambda =10^{-3}$ for the model where the burst rate
is constant per galaxy, $\Lambda =3 \times 10^{-4}$ when the burst
rate is proportional to the galaxy mass, $\Lambda =10^{-2}$ when
the burst rate is proportional to the galaxy luminosity, and
$\Lambda =10^{-4}$ for the model where the burst rate is
proportional to the product of the galaxy mass and the cosmic star
formation rate. Thus the third model is preferred; we will use the
corresponding $p_B$ in adding the four bursts with only host
galaxy R magnitudes to the sample with spectroscopic redshifts.

\section{The Burst Energy Distributions}

\begin{figure}
\plotone{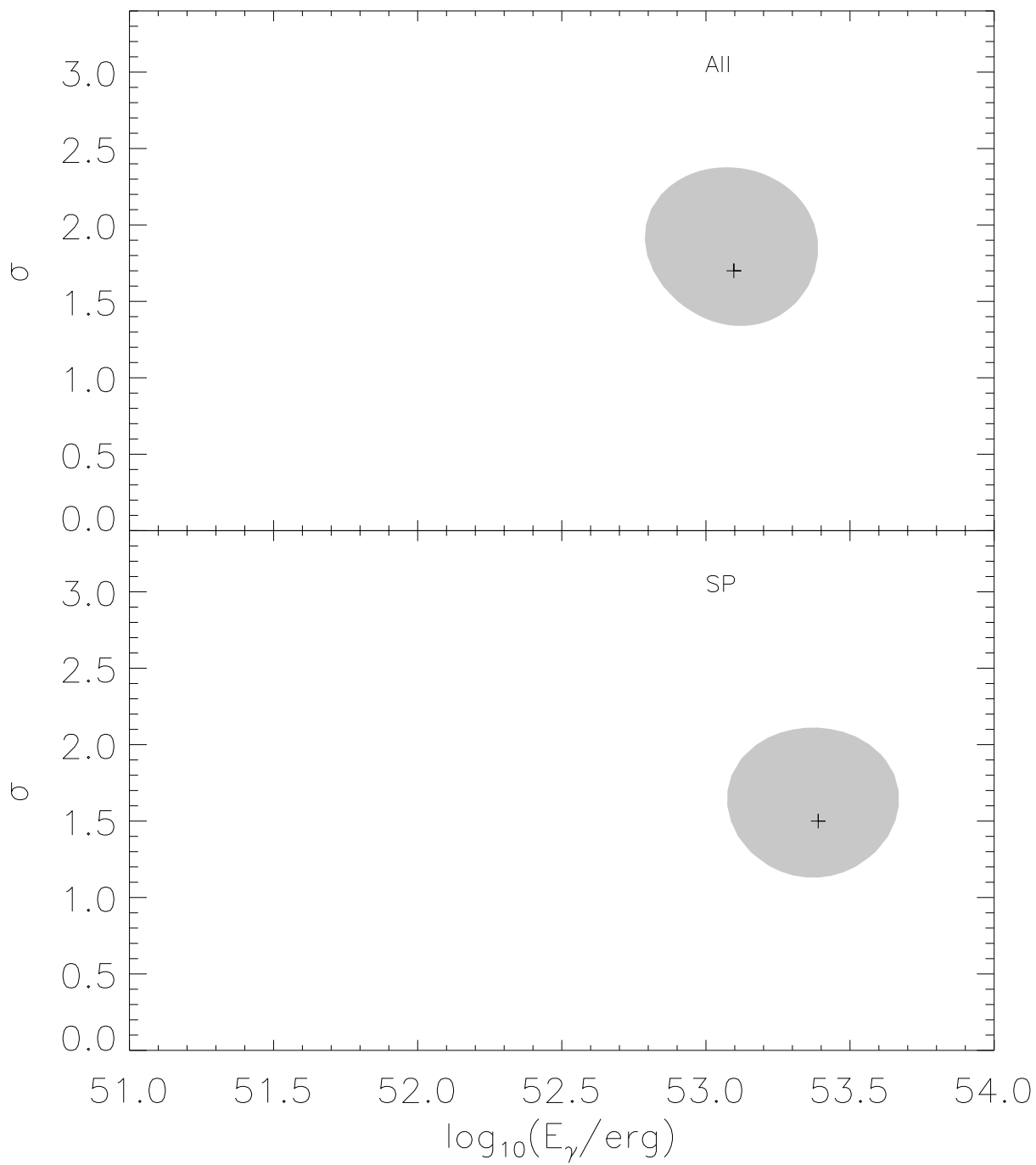}
\caption{
The 70\% confidence likelihood contour plot for
total burst energy $\langle E_{\gamma iso} \rangle$ and its
variance $\sigma_\gamma$ (see text) in the 20--2000 keV band when
using the 8 bursts with spectroscopic redshifts (top panel) and
when adding to these 8 bursts the 4 GRBs with only host galaxy
magnitudes (bottom panel).\label{tot}}
\end{figure}

We first estimate the characteristic isotropic gamma-ray energy ($\langle
E_{\gamma iso} \rangle$) and its spread ($\sigma_\gamma$) in the 20--2000 keV
band. The result is presented in Figure~\ref{tot}. The upper panel shows the
likelihood contour corresponding to 70\% confidence using only the 8 GRBs with
spectroscopic redshifts, while the bottom panel shows the same contour level
but with the additional 4 GRBs with statistical redshifts. Therefore, the
preferred value for $\langle E_{\gamma iso} \rangle$ is $1.3^{+1.2}_{-1.0}
\times 10^{53}$ erg with $\sigma_\gamma=1.7^{+0.7}_{-0.3}$. Note that there is
no need to k-correct the $\gamma$-ray data in this case because the 20--2000
keV band already contains most of the energy of the GRB.

\begin{figure}
\plotone{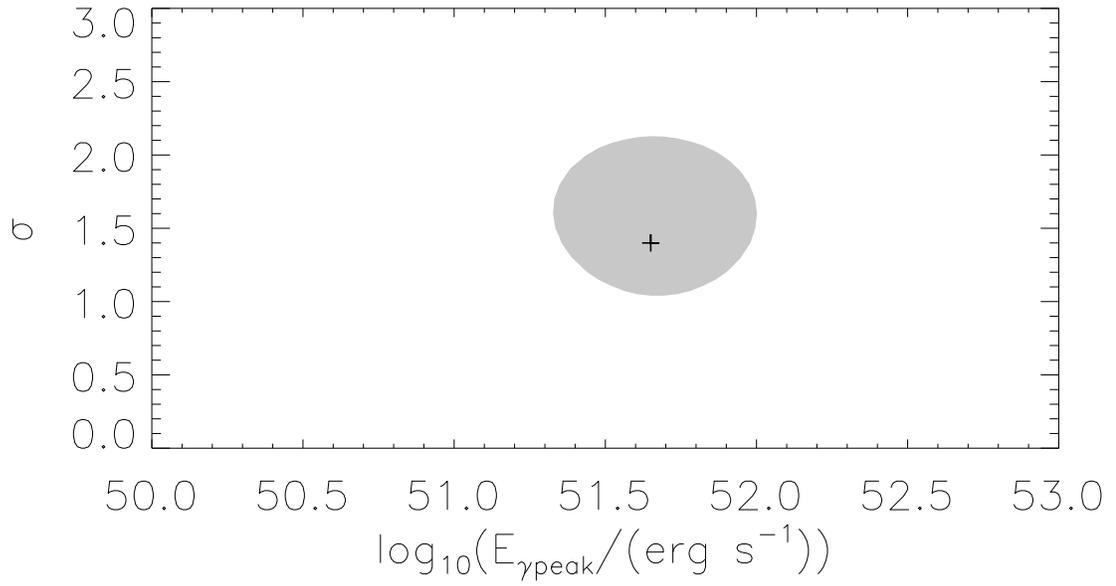}
\caption{
The 70\% confidence likelihood contour plot for
the peak burst luminosity $\langle L_{\gamma iso} \rangle$ and
$\sigma_L$ in the 50--300 keV band. Note that in this case we have
k-corrected the data from Table~\ref{Table} using the spectral
fits provided in the table.\label{likpeak}}
\end{figure}

Similarly we derive the peak luminosity $\langle L_{\gamma iso} \rangle$ in
the 50--300 keV band. In this case we apply a k-correction to the data using
the spectral fits described in Table~\ref{Table}. Figure~\ref{likpeak} shows
the 70\% confidence contour plot using all the GRBs in our sample. We find
$\langle L_{\gamma iso} \rangle=4.6^{+5.4}_{-2.5} \times 10^{51}$ erg s$^{-1}$
with $\sigma_L=1.4^{+0.7}_{-0.4}$.

\begin{figure}
\plotone{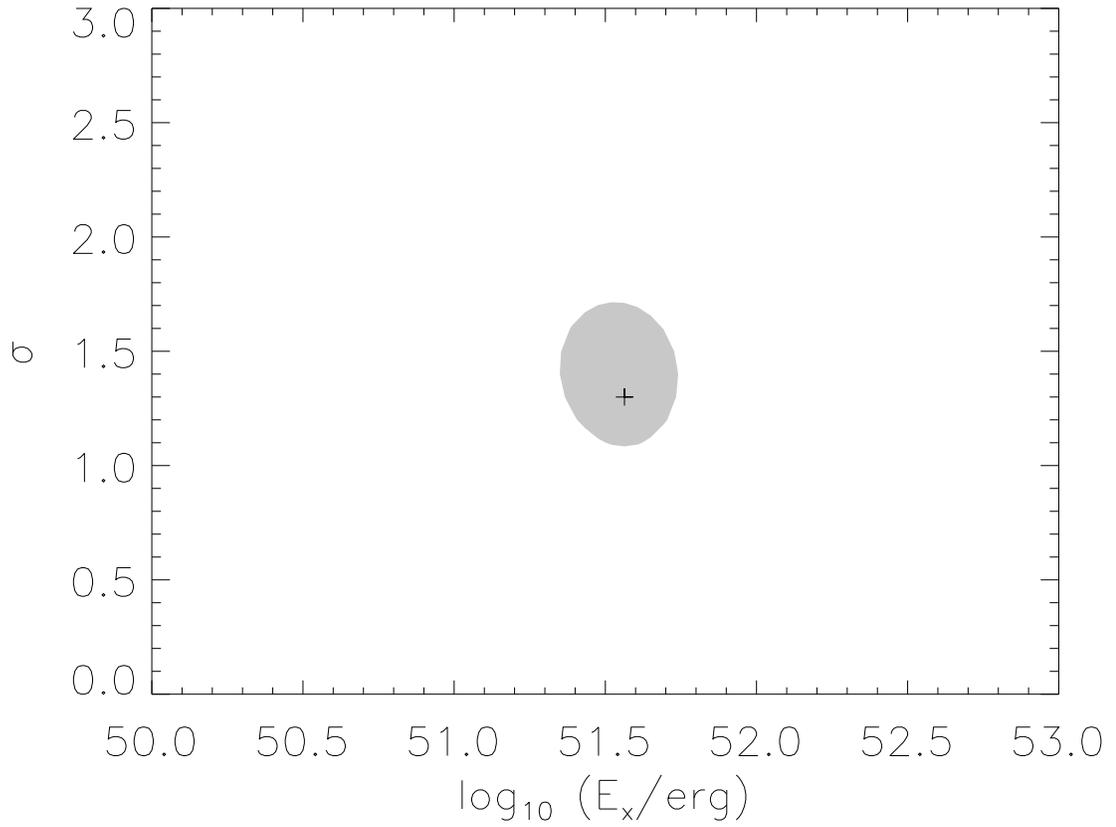}
\caption{
The 70\% confidence likelihood contour plot for
the total afterglow X-ray energy $\langle E_{X iso} \rangle$ and
$\sigma_X$. We integrated the X-ray flux between 100 and $10^5$
seconds after the burst in the 2--10 keV band and applied a
corresponding k-correction assuming a spectrum $\propto
\nu^{-0.75}$.\label{xafter}}  
\end{figure}

Finally, we find the average energy of the X-ray afterglow.  To homogenize the
observations, we integrate the X-ray emission between 100 and 10$^5$ seconds
after the burst using the best fitting power law to the observed light curve
(slopes for the different light curves are given in Table~\ref{Table}).
Figure~\ref{xafter} shows the $1\sigma$ contour plot when the 7 GRBs with
X-ray light curves are considered. Here we find that $\langle E_{X iso}
\rangle = 4.0^{+1.6}_{-1.8} \times 10^{51}$ erg and $\sigma_X =
1.3^{+0.4}_{-0.3}$.

\section{Discussion and Conclusions}

We have determined the distributions of the total burst energy,
the peak burst luminosity and the total X-ray afterglow energy for
a sample of bursts with either spectroscopic redshifts or host
galaxy R magnitudes. These distributions reflect the physics of
the burst process.

We have found that the best model for the burst rate is
proportional to the host galaxy luminosity and not to the host
size. This result provides further support to the proposition that
the bursts are associated with star formation. Star forming
galaxies are brighter than other galaxies with the same size.

It seems that $\langle E_{\gamma iso} \rangle$ is slightly broader than
$\langle E_{X iso} \rangle$, but within the 70\% confidence both distributions
are compatible with having the same width.  However, inspection of the data
points shows that X-ray energy of one burst---GRB970508---significantly
broadens the X-ray distribution; otherwise the X-ray distribution would be
{\it much} narrower than the gamma-ray distribution.  Thus we must conclude
that the data are insufficient to determine conclusively the width of the
luminosity functions, although there is a trend for the X-ray distribution to
be narrower than the ${\gamma}$ distribution.

The ratio between the widths of the gamma-ray and the X-ray luminosity
functions is particularly important as far as the physics of the physical
processes within the ``inner engine.'' A ratio of order unity suggests that
the ``inner engine'' emits a rather uniform flow with no significant angular
variation on an angular scale of few degrees (corresponding to the width seen
several hours after the burst while the X-rays are emitted).  On the other
hand, a large $\sigma_\gamma / \sigma_X$ would support the ``patchy shell''
model\citep{KP99} which predicts that the gamma-ray energy emitted during the
GRB has a significantly wider distribution than the X-ray energy emitted
during the afterglow, $\sigma_\gamma>\sigma_X$.  In this model, there are mass
fluctuations on the shells whose collisions produce the internal shocks which
subsequently radiate the observed gamma-rays. Thus the gamma-ray intensity
should vary greatly with the observer's angle. On the other hand, the
afterglow results from the decelerating external shock, and the observer sees
a much larger solid angle of the relativistic outflow in the afterglow than in
the burst itself. Thus there are smaller fluctuations in the afterglow
intensity with observer angle. Finally $\sigma_\gamma / \sigma_X < 1$ would be
indicative of extremely large variations (from one burst to the other) in the
circumstellar matter surrounding the GRBs.  The data are sufficient to rule
out this third possibility, but cannot distinguish with confidence between the
first and second, although the trend is toward favoring the ``patchy shell''
model.

It is interesting to compare our results with the energy distribution
determined for the whole GRB sample. \cite{S99} finds for a broken
power law distribution that the energy of the break is $1.2 \times 10^{53}$
ergs, which is very close to our average energy. However, the average energy
in Schmidt's distribution is a factor of $\sim5$ smaller than this break
energy. This can be explained by the fact that afterglow can be observed only
for the more luminous bursts for which we can determine an exact position.

Finally we note that we have presented here a new methodology for obtaining
the luminosity function of a sample for which the magnitude is known but there
are no redshift measurements. This method can be best applied to other
populations that like GRBs have an intrinsically wide luminosity function.
While a galaxy's optical magnitude does not provide a reliable distance
measure for a single object, is should be sufficiently reliable when analyzing
a large sample. This method should be tested on other samples in the future.

This methodology can be used wherever burst distances and energies are
required. For example, a burst sample with known distances is required to
calibrate proposed correlations between the burst energy and the
frequency--dependent lags in pulses \citep{FR2001} or light-curve variability
\citep{NMB2000}.

This research was supported in part by a US-Israel BSF.  The work of David
Band was performed under the auspices of the U.S.  Department of Energy by the
Los Alamos National Laboratory under Contract No. W-7405-Eng-36.

\clearpage


\clearpage
\begin{deluxetable}{l l l l l l l l l l l l l}
\tablecolumns{13}
\tabletypesize{\tiny}
\tablewidth{0pt}
\rotate
\tablecaption{\label{Table}Main properties of the sample of Gamma Ray Bursts.}
\tablehead{
\colhead{Name}
&\colhead{$N_0$}
&\colhead{$\alpha$}
&\colhead{$\beta$}
&\colhead{$E_0$ (keV)}
&\colhead{$F_{\gamma}$ (erg cm$^{-2}$)\tablenotemark{a}}
&\colhead{$F_{\gamma peak}$ (erg cm$^{-2}$ s$^{-1}$) \tablenotemark{b}}
&\colhead{$\delta_x$}
&\colhead{$F_x$ (erg cm$^{-2}$ s$^{-1}$)}
&\colhead{$t$ (s)}
&\colhead{$R_{\rm gal}$}
&\colhead{$z_{\rm obs}$}
&\colhead{$z_{\rm est}$}
}
\startdata
970111& $7.86\times 10^{-2}$ & $-0.533$ & \nodata &106.56& $4.07\times
10^{-5}$ &
   \nodata & \nodata & \nodata &35.36& \nodata & \nodata & \nodata\\
%
970228& \nodata & \nodata & \nodata & \nodata & \nodata & $2.15\times
10^{-6}$ &
   \nodata & \nodata & \nodata & 24.6&0.695&0.8\\
%
970508& $1.73\times 10^{-3}$ & $-1.191$ & $-1.831$ &480.84& $5.54\times
10^{-6}$ & $7.38\times 10^{-7}$ & $-1.2$ & $3.9\times 10^{-7}$ & 23.44&
25.8&0.835&1.2\\
970616& $4.99\times 10^{-3}$ & $-1.464$ & $-2.549$ & 296.91&
   $4.03\times 10^{-5}$ & \nodata &
   \nodata & \nodata & 203.68& \nodata & \nodata & \nodata\\
970815& $4.17\times 10^{-3}$ & $-1.094$ & $-2.802$ &113.25& $1.37\times
10^{-5}$ &
   \nodata & \nodata & \nodata &183.59& \nodata & \nodata & \nodata\\
%
%
970828& $1.15\times 10^{-2}$ & $-0.704$ & $-2.072$ & 229.74& $9.60\times
10^{-5}$ & $3.01\times 10^{-6}$ & $-1.0$ & $1.82\times 10^{-6}$ &
146.59& 24.5&0.958&0.8 \\
971024& $9.72\times 10^{-3}$ &0.185& $-3.881$ &40.93& $2.38\times 10^{-6}$ &
   \nodata & \nodata & \nodata &97.82& \nodata & \nodata & \nodata\\
%
971214& $7.23\times 10^{-3}$ & $-0.783$ & $-2.574$ &155.96& $9.44\times
10^{-6}$ &
   $2.28\times 10^{-6}$ & $-1.6$ & $1.31\times 10^{-6}$ &45.45& 26.2
&3.412&1.2\\
971227& $1.06\times 10^{-2}$ & $-1.440$ & $-4.198$ &112.03& $1.21\times
10^{-6}$ & $3.37\times 10^{-6}$ & \nodata & \nodata &6.94& 25.0& \nodata &1.2\\
980109& $9.34\times 10^{-3}$ & $-0.428$ & $-2.291$ &62.65&
$4.08\times 10^{-6}$ & \nodata &
   \nodata & \nodata &42.97& \nodata & \nodata & \nodata\\
980326& $2.11\times 10^{-2}$ & $-1.327$ & $-4.335$ &77.19& $9.22\times
10^{-7}$ &
   $8.00\times 10^{-7}$ & \nodata & \nodata &4.01& 25.3& \nodata &1.2\\
980329 & $2.58\times 10^{-2}$ & $-0.964$ & $-2.431$ & 235.65& $5.51\times
10^{-5}$ & $1.15\times 10^{-5}$ & $-1.34$ & $0.56\times 10^{-6}$
&50.15& 26.3& \nodata &1.2\\
%
980425& $4.70\times 10^{-3}$ & $-1.266$ & \nodata &161.20& $3.87\times
10^{-6}$ &
   \nodata & \nodata & \nodata &37.41&14.3&0.0085&0.01\\
%
980519& $4.79\times 10^{-3}$ & $-1.352$ & \nodata &315.94& $1.03\times
10^{-5}$ & $2.98\times 10^{-6}$ & \nodata & \nodata &56.35& 24.7& \nodata
&0.8\\
%
980613& \nodata & \nodata & \nodata & \nodata & \nodata & $3.87\times
10^{-7}$ & $-1.6$ & $0.55\times 10^{-6}$ & \nodata & 23.85&1.096&0.8 \\
%
980703& $4.41\times 10^{-3}$ & $-1.314$ & $-2.396$ &370.26& $2.26\times
10^{-5}$ & $1.62\times 10^{-6}$ & $-1.67$ & $1.27\times
10^{-6}$ &102.37& 22.8 &0.966&0.6\\
980706& $1.50\times 10^{-3}$ & $-1.112$ & \nodata & $5000.0$ & $2.31\times
10^{-5}$ & \nodata & \nodata & \nodata &72.81& \nodata & \nodata & \nodata\\
%
990123& $2.62\times 10^{-2}$ & $-0.900$ & $-2.476$ &549.51& $2.68\times
10^{-4}$ & $1.11\times 10^{-5}$ & $-1.6$ & $8.98\times
10^{-6}$ &104.61& 24.3&1.600&0.8\\
990506 & $1.51\times 10^{-2}$ & $-1.370$ & $-2.152$ &449.78& $1.94\times
10^{-4}$ & \nodata & \nodata & \nodata & 220.38& 25.0 &1.2& \nodata\\
990510 & $7.96\times 10^{-3}$ & $-1.275$ & $-2.670$ &174.24& $2.26\times
10^{-5}$ & $5.02\times
10^{-6}$ & \nodata & \nodata &103.84& \nodata &1.619& \nodata\\
990712& \nodata & \nodata & \nodata & \nodata & \nodata & \nodata &
\nodata & \nodata & \nodata & 22.0&0.43&0.40\\
990806& $8.91\times 10^{-3}$ & $-0.658$ & $-2.261$ &109.09& $2.51\times
10^{-6}$ & \nodata & \nodata & \nodata &12.71& \nodata & \nodata & \nodata\\
991014& $1.16\times 10^{-2}$ & $-0.754$ & $-2.190$ &84.65& $1.01\times
10^{-6}$ & \nodata & \nodata & \nodata &5.093& \nodata & \nodata & \nodata\\
991105& $2.16\times 10^{-3}$ & $-1.734$ & \nodata &363.49& $2.80\times
10^{-6}$ & \nodata & \nodata & \nodata &34.28& \nodata & \nodata & \nodata\\
991208 & \nodata & \nodata & \nodata & \nodata & \nodata & \nodata & \nodata
& \nodata & \nodata & \nodata & 0.706 & \nodata \\
991216& $1.29\times 10^{-1}$ & $-1.234$ & $-2.184$ &414.83& $1.94\times
10^{-4}$ & \nodata & \nodata & \nodata & 24.96& 26.9 & 1.02 & \nodata\\
991229 & $5.76\times10^{-3}$ & $-1.171$ & \nodata & 3000 &
$1.70\times10^{-4}$ & \nodata & \nodata & \nodata & 174.431 &
\nodata & \nodata & \nodata\\
000115 & $5.92\times 10^{-2}$ & $-0.856$ & $-2.692$ & 191.74 &
$3.68\times 10^{-5}$ & \nodata & \nodata & \nodata &
18.214& \nodata & \nodata & \nodata\\
000126 & $7.71\times 10^{-3}$ & $-1.279$ & $-2.289$ &342.64 &
$3.28\times 10^{-5}$ & \nodata & \nodata & \nodata &
85.023& \nodata & \nodata & \nodata\\
000131 & $1.68\times 10^{-2}$ & $-0.688$ & $-2.068$ & 98.982 &
$4.18\times 10^{-5}$ & \nodata & \nodata & \nodata &
110.151& \nodata & \nodata & \nodata\\
000201 & $5.80\times10^{-3}$ & $-1.092$ & $-5.000$ & 347.73 &
$2.81\times 10^{-5}$ & \nodata & \nodata & \nodata &
99.359& \nodata & \nodata & \nodata\\
000301A & $3.01\times 10^{-3}$ & $-0.901$ & $-2.347$ & 468.48 &
$6.17\times 10^{-6}$ & \nodata & \nodata & \nodata &
23.757& \nodata & \nodata & \nodata\\
%
0000301C & \nodata & \nodata & \nodata & \nodata & \nodata & \nodata & \nodata
& \nodata & \nodata & \nodata & 2.0335 & \nodata \\
000307 & $1.05\times10^{-2}$ & $-1.372$ & $-2.267$ & 168.63 &
$8.70\times 10^{-6}$ & \nodata & \nodata & \nodata &
26.598& \nodata & \nodata & \nodata\\
000408 & $5.48\times 10^{-2}$ & $-1.208$ & $-2.353$ & 248.16 &
$2.28\times 10^{-5}$ & \nodata & \nodata & \nodata &
10.278& \nodata & \nodata & \nodata\\
%
000418 & \nodata & \nodata & \nodata & \nodata &
\nodata & \nodata & \nodata & \nodata &
\nodata & 23.9 & 1.118 & \nodata\\
000429 & $2.37\times 10^{-3}$ & $-1.050$ & $-5.000$ & 536.80 &
$3.25\times 10^{-5}$ & \nodata & \nodata & \nodata &
179.693& \nodata & \nodata & \nodata\\
000508B & $1.18\times 10^{-2}$ & $-0.711$ & $-2.247$ & 105.48 &
$1.49\times 10^{-5}$ & \nodata & \nodata & \nodata &
59.283& \nodata & \nodata & \nodata\\
000519 & $8.90\times 10^{-3}$ & $-1.500$ & $-2.361$ & 473.81 &
$7.94\times 10^{-6}$ & \nodata & \nodata & \nodata &
16.991& \nodata & \nodata & \nodata\\
\enddata

\tablenotetext{a}{20--2000 keV}
\tablenotetext{b}{50--300 keV}

\end{deluxetable}

\clearpage

\clearpage

\clearpage

\clearpage

\clearpage

\end{document}